\catcode`@=11 
%
%
%
%

\font\fourteenrm=cmr10 scaled\magstep2
\font\twelverm=cmr12

\font\ninerm=cmr9

\font\sixrm=cmr6

\font\fourteenbf=cmbx10 scaled\magstep2
\font\twelvebf=cmbx10 scaled\magstep1
\font\ninebf=cmbx9	      \font\sixbf=cmbx6
\font\seventeeni=cmmi10 scaled\magstep3	    \skewchar\seventeeni='177
\font\fourteeni=cmmi10 scaled\magstep2	    \skewchar\fourteeni='177
\font\twelvei=cmmi10 scaled\magstep1	    \skewchar\twelvei='177
\font\ninei=cmmi9			    \skewchar\ninei='177
\font\sixi=cmmi6			    \skewchar\sixi='177
\font\seventeensy=cmsy10 scaled\magstep3    \skewchar\seventeensy='60
\font\fourteensy=cmsy10 scaled\magstep2	    \skewchar\fourteensy='60
\font\twelvesy=cmsy10 scaled\magstep1	    \skewchar\twelvesy='60
\font\ninesy=cmsy9			    \skewchar\ninesy='60
\font\sixsy=cmsy6			    \skewchar\sixsy='60

\font\fourteenex=cmex10 scaled\magstep2
\font\twelveex=cmex10 scaled\magstep1

\font\fourteensl=cmsl10 scaled\magstep2
\font\twelvesl=cmsl10 scaled\magstep1

\font\ninesl=cmsl9

\font\fourteenit=cmti10 scaled\magstep2
\font\twelveit=cmti10 scaled\magstep1
\font\twelvett=cmtt10 scaled\magstep1

%

\font\fourteencp=cmcsc10 scaled\magstep2
\font\twelvecp=cmcsc10 scaled\magstep1
\font\tencp=cmcsc10
\newfam\cpfam
%
%
\newcount\f@ntkey	     \f@ntkey=0
\def\samef@nt{\relax \ifcase\f@ntkey \rm \or\oldstyle \or\or
	 \or\it \or\sl \or\bf \or\tt \or\caps \fi }
\def\fourteenpoint{\relax
    \textfont0=\fourteenrm	    \scriptfont0=\tenrm
    \scriptscriptfont0=\sevenrm
     \def\rm{\fam0 \fourteenrm \f@ntkey=0 }\relax
    \textfont1=\fourteeni	    \scriptfont1=\teni
    \scriptscriptfont1=\seveni
     \def\oldstyle{\fam1 \fourteeni\f@ntkey=1 }\relax
    \textfont2=\fourteensy	    \scriptfont2=\tensy
    \scriptscriptfont2=\sevensy
    \textfont3=\fourteenex     \scriptfont3=\fourteenex
    \scriptscriptfont3=\fourteenex
    \def\it{\fam\itfam \fourteenit\f@ntkey=4 }\textfont\itfam=\fourteenit
    \def\sl{\fam\slfam \fourteensl\f@ntkey=5 }\textfont\slfam=\fourteensl
    \scriptfont\slfam=\tensl
    \def\bf{\fam\bffam \fourteenbf\f@ntkey=6 }\textfont\bffam=\fourteenbf
    \scriptfont\bffam=\tenbf	 \scriptscriptfont\bffam=\sevenbf
    \def\tt{\fam\ttfam \twelvett \f@ntkey=7 }\textfont\ttfam=\twelvett
    \h@big=11.9\p@{} \h@Big=16.1\p@{} \h@bigg=20.3\p@{} \h@Bigg=24.5\p@{}
    \def\caps{\fam\cpfam \twelvecp \f@ntkey=8 }\textfont\cpfam=\twelvecp
    \setbox\strutbox=\hbox{\vrule height 12pt depth 5pt width\z@}
    \samef@nt}
\def\twelvepoint{\relax
    \textfont0=\twelverm	  \scriptfont0=\ninerm
    \scriptscriptfont0=\sixrm
     \def\rm{\fam0 \twelverm \f@ntkey=0 }\relax
    \textfont1=\twelvei		  \scriptfont1=\ninei
    \scriptscriptfont1=\sixi
     \def\oldstyle{\fam1 \twelvei\f@ntkey=1 }\relax
    \textfont2=\twelvesy	  \scriptfont2=\ninesy
    \scriptscriptfont2=\sixsy
    \textfont3=\twelveex	  \scriptfont3=\twelveex
    \scriptscriptfont3=\twelveex
    \def\it{\fam\itfam \twelveit \f@ntkey=4 }\textfont\itfam=\twelveit
    \def\sl{\fam\slfam \twelvesl \f@ntkey=5 }\textfont\slfam=\twelvesl
    \scriptfont\slfam=\ninesl
    \def\bf{\fam\bffam \twelvebf \f@ntkey=6 }\textfont\bffam=\twelvebf
    \scriptfont\bffam=\ninebf	  \scriptscriptfont\bffam=\sixbf
    \def\tt{\fam\ttfam \twelvett \f@ntkey=7 }\textfont\ttfam=\twelvett
    \h@big=10.2\p@{}
    \h@Big=13.8\p@{}
    \h@bigg=17.4\p@{}
    \h@Bigg=21.0\p@{}
    \def\caps{\fam\cpfam \twelvecp \f@ntkey=8 }\textfont\cpfam=\twelvecp
    \setbox\strutbox=\hbox{\vrule height 10pt depth 4pt width\z@}
    \samef@nt}
\def\tenpoint{\relax
    \textfont0=\tenrm	       \scriptfont0=\sevenrm
    \scriptscriptfont0=\fiverm
    \def\rm{\fam0 \tenrm \f@ntkey=0 }\relax
    \textfont1=\teni	       \scriptfont1=\seveni
    \scriptscriptfont1=\fivei
    \def\oldstyle{\fam1 \teni \f@ntkey=1 }\relax
    \textfont2=\tensy	       \scriptfont2=\sevensy
    \scriptscriptfont2=\fivesy
    \textfont3=\tenex	       \scriptfont3=\tenex
    \scriptscriptfont3=\tenex
    \def\it{\fam\itfam \tenit \f@ntkey=4 }\textfont\itfam=\tenit
    \def\sl{\fam\slfam \tensl \f@ntkey=5 }\textfont\slfam=\tensl
    \def\bf{\fam\bffam \tenbf \f@ntkey=6 }\textfont\bffam=\tenbf
    \scriptfont\bffam=\sevenbf	   \scriptscriptfont\bffam=\fivebf
    \def\tt{\fam\ttfam \tentt \f@ntkey=7 }\textfont\ttfam=\tentt
    \def\caps{\fam\cpfam \tencp \f@ntkey=8 }\textfont\cpfam=\tencp
    \setbox\strutbox=\hbox{\vrule height 8.5pt depth 3.5pt width\z@}
    \samef@nt}
%
%
%
%
\newdimen\h@big  \h@big=8.5\p@
\newdimen\h@Big  \h@Big=11.5\p@
\newdimen\h@bigg  \h@bigg=14.5\p@
\newdimen\h@Bigg  \h@Bigg=17.5\p@
\def\big#1{{\hbox{$\left#1\vbox to\h@big{}\right.\n@space$}}}
\def\Big#1{{\hbox{$\left#1\vbox to\h@Big{}\right.\n@space$}}}
\def\bigg#1{{\hbox{$\left#1\vbox to\h@bigg{}\right.\n@space$}}}
\def\Bigg#1{{\hbox{$\left#1\vbox to\h@Bigg{}\right.\n@space$}}}
%
%
%

\normalbaselineskip = 18.0pt plus 0.2pt minus 0.1pt
\normallineskip = 1.5pt plus 0.1pt minus 0.1pt
\normallineskiplimit = 1.5pt
\newskip\normaldisplayskip
\normaldisplayskip = 20pt plus 5pt minus 10pt
\newskip\normaldispshortskip
\normaldispshortskip = 6pt plus 5pt
\newskip\normalparskip
\normalparskip = 6pt plus 2pt minus 1pt
\newskip\skipregister
\skipregister = 5pt plus 2pt minus 1.5pt
\newif\ifsingl@	   \newif\ifdoubl@
\newif\iftwelv@	   \twelv@true
\def\singlespace{\singl@true\doubl@false\spaces@t}
\def\doublespace{\singl@false\doubl@true\spaces@t}
\def\normalspace{\singl@false\doubl@false\spaces@t}
\def\Tenpoint{\tenpoint\twelv@false\spaces@t}
\def\Twelvepoint{\twelvepoint\twelv@true\spaces@t}
\def\spaces@t{\relax%
 \iftwelv@ \ifsingl@\subspaces@t3:4;\else\subspaces@t1:1;\fi%
 \else \ifsingl@\subspaces@t3:5;\else\subspaces@t4:5;\fi \fi%
 \ifdoubl@ \multiply\baselineskip by 5%
 \divide\baselineskip by 4 \fi \unskip}
\def\subspaces@t#1:#2;{%
      \baselineskip = \normalbaselineskip%
      \multiply\baselineskip by #1 \divide\baselineskip by #2%
      \lineskip = \normallineskip%
      \multiply\lineskip by #1 \divide\lineskip by #2%
      \lineskiplimit = \normallineskiplimit%
      \multiply\lineskiplimit by #1 \divide\lineskiplimit by #2%
      \parskip = \normalparskip%
      \multiply\parskip by #1 \divide\parskip by #2%
      \abovedisplayskip = \normaldisplayskip%
      \multiply\abovedisplayskip by #1 \divide\abovedisplayskip by #2%
      \belowdisplayskip = \abovedisplayskip%
      \abovedisplayshortskip = \normaldispshortskip%
      \multiply\abovedisplayshortskip by #1%
	\divide\abovedisplayshortskip by #2%
      \belowdisplayshortskip = \abovedisplayshortskip%
      \advance\belowdisplayshortskip by \belowdisplayskip%
      \divide\belowdisplayshortskip by 2%
      \smallskipamount = \skipregister%
      \multiply\smallskipamount by #1 \divide\smallskipamount by #2%
      \medskipamount = \smallskipamount \multiply\medskipamount by 2%
      \bigskipamount = \smallskipamount \multiply\bigskipamount by 4 }
\def\normalbaselines{ \baselineskip=\normalbaselineskip%
   \lineskip=\normallineskip \lineskiplimit=\normallineskip%
   \iftwelv@\else \multiply\baselineskip by 4 \divide\baselineskip by 5%
     \multiply\lineskiplimit by 4 \divide\lineskiplimit by 5%
     \multiply\lineskip by 4 \divide\lineskip by 5 \fi }
\Twelvepoint  
\interlinepenalty=50
\interfootnotelinepenalty=5000
\predisplaypenalty=9000
\postdisplaypenalty=500
\hfuzz=1pt
\vfuzz=0.2pt
%
%
%
\def\pagecontents{%
   \ifvoid\topins\else\unvbox\topins\vskip\skip\topins\fi
   \dimen@ = \dp255 \unvbox255
   \ifvoid\footins\else\vskip\skip\footins\footrule\unvbox\footins\fi
   \ifr@ggedbottom \kern-\dimen@ \vfil \fi }
\def\makeheadline{\vbox to 0pt{ \skip@=\topskip
      \advance\skip@ by -12pt \advance\skip@ by -2\normalbaselineskip
      \vskip\skip@ \line{\vbox to 12pt{}\the\headline} \vss
      }\nointerlineskip}
\def\makefootline{\baselineskip = 1.5\normalbaselineskip
		 \line{\the\footline}}
\newif\iffrontpage
\newif\ifletterstyle
\newif\ifp@genum
\def\nopagenumbers{\p@genumfalse}
\def\pagenumbers{\p@genumtrue}
\pagenumbers
\newtoks\paperheadline
\newtoks\letterheadline
\newtoks\letterfrontheadline
\newtoks\lettermainheadline
\newtoks\paperfootline
\newtoks\letterfootline
\newtoks\date
\footline={\ifletterstyle\the\letterfootline\else\the\paperfootline\fi}
\paperfootline={\hss\iffrontpage\else\ifp@genum\tenrm
    -- \folio\ --\hss\fi\fi}
\letterfootline={\hfil}
\headline={\ifletterstyle\the\letterheadline\else\the\paperheadline\fi}
\paperheadline={\hfil}
\letterheadline{\iffrontpage\the\letterfrontheadline
     \else\the\lettermainheadline\fi}
\lettermainheadline={\rm\ifp@genum page \ \folio\fi\hfil\the\date}
\def\monthname{\relax\ifcase\month 0/\or January\or February\or
   March\or April\or May\or June\or July\or August\or September\or
   October\or November\or December\else\number\month/\fi}
\date={\monthname\ \number\day, \number\year}
\countdef\pagenumber=1  \pagenumber=1
\def\advancepageno{\global\advance\pageno by 1
   \ifnum\pagenumber<0 \global\advance\pagenumber by -1
    \else\global\advance\pagenumber by 1 \fi \global\frontpagefalse }
\def\folio{\ifnum\pagenumber<0 \romannumeral-\pagenumber
	   \else \number\pagenumber \fi }
\def\footrule{\dimen@=\prevdepth\nointerlineskip
   \vbox to 0pt{\vskip -0.25\baselineskip \hrule width 0.35\hsize \vss}
   \prevdepth=\dimen@ }
\newtoks\foottokens
\foottokens={\Tenpoint\singlespace}
\newdimen\footindent
\footindent=24pt
\def\vfootnote#1{\insert\footins\bgroup  \the\foottokens
   \interlinepenalty=\interfootnotelinepenalty \floatingpenalty=20000
   \splittopskip=\ht\strutbox \boxmaxdepth=\dp\strutbox
   \leftskip=\footindent \rightskip=\z@skip
   \parindent=0.5\footindent \parfillskip=0pt plus 1fil
   \spaceskip=\z@skip \xspaceskip=\z@skip
   \Textindent{$ #1 $}\footstrut\futurelet\next\fo@t}
\def\Textindent#1{\noindent\llap{#1\enspace}\ignorespaces}
\def\footnote#1{\attach{#1}\vfootnote{#1}}

\def\foot{\attach\footsymbolgen\vfootnote{\footsymbol}}
\let\footsymbol=\star
\newcount\lastf@@t	     \lastf@@t=-1
\newcount\footsymbolcount    \footsymbolcount=0
\newif\ifPhysRev
\def\footsymbolgen{\relax \ifPhysRev \iffrontpage \NPsymbolgen\else
      \PRsymbolgen\fi \else \NPsymbolgen\fi
   \global\lastf@@t=\pageno \footsymbol }
\def\NPsymbolgen{\ifnum\footsymbolcount<0 \global\footsymbolcount=0\fi
   {\iffrontpage \else \advance\lastf@@t by 1 \fi
    \ifnum\lastf@@t<\pageno \global\footsymbolcount=0
     \else \global\advance\footsymbolcount by 1 \fi }
   \ifcase\footsymbolcount \fd@f\star\or \fd@f\dagger\or \fd@f\ast\or
    \fd@f\ddagger\or \fd@f\natural\or \fd@f\diamond\or \fd@f\bullet\or
    \fd@f\nabla\else \fd@f\dagger\global\footsymbolcount=0 \fi }
\def\fd@f#1{\xdef\footsymbol{#1}}
\def\PRsymbolgen{\ifnum\footsymbolcount>0 \global\footsymbolcount=0\fi
      \global\advance\footsymbolcount by -1
      \xdef\footsymbol{\sharp\number-\footsymbolcount} }
\def\space@ver#1{\let\@sf=\empty \ifmmode #1\else \ifhmode
   \edef\@sf{\spacefactor=\the\spacefactor}\unskip${}#1$\relax\fi\fi}
\def\attach#1{\space@ver{\strut^{\mkern 2mu #1} }\@sf\ }
%
%
%
\newcount\chapternumber	     \chapternumber=0
\newcount\sectionnumber	     \sectionnumber=0
\newcount\equanumber	     \equanumber=0
\let\chapterlabel=0
\newtoks\chapterstyle	     \chapterstyle={\Number}
\newskip\chapterskip	     \chapterskip=\bigskipamount
\newskip\sectionskip	     \sectionskip=\medskipamount
\newskip\headskip	     \headskip=8pt plus 3pt minus 3pt
\newdimen\chapterminspace    \chapterminspace=15pc
\newdimen\sectionminspace    \sectionminspace=10pc
\newdimen\referenceminspace  \referenceminspace=25pc
\def\chapterreset{\global\advance\chapternumber by 1
   \ifnum\equanumber<0 \else\global\equanumber=0\fi
   \sectionnumber=0 \makel@bel}
\def\makel@bel{\xdef\chapterlabel{
  {\ifPhysRev \Roman{\the\chapternumber}. \hskip-4pt
         \else \Number{\the\chapternumber}. \hskip -4pt \fi}}}
\newtoks\cstyle       \cstyle={\Number}
\def\clabel{\the\cstyle{\the\chapternumber}.}
\def\sectionlabel{\ifPhysRev {\Alphabetic\sectionnumber.} \else
      {\number\sectionnumber.} \fi}
\def\alphabetic#1{\count255='140 \advance\count255 by #1\char\count255}
\def\Alphabetic#1{\count255='100 \advance\count255 by #1\char\count255}
\def\Roman#1{\uppercase\expandafter{\romannumeral #1}}
\def\roman#1{\romannumeral #1}
\def\Number#1{\number #1}
\def\unnumberedchapters{\let\makel@bel=\relax \let\chapterlabel=\relax
\let\sectionlabel=\relax \equanumber=-1 }
%
\def\titlestyle#1{\par\begingroup \interlinepenalty=9999
     \leftskip=0.03\hsize plus 0.20\hsize minus 0.03\hsize
     \rightskip=\leftskip \parfillskip=0pt
     \hyphenpenalty=9000 \exhyphenpenalty=9000
     \tolerance=9999 \pretolerance=9000
     \spaceskip=0.333em \xspaceskip=0.5em
     \iftwelv@\fourteenpoint\fourteenbf\else\twelvepoint\twelvebf\fi
     \noindent  #1\par\endgroup }
\def\spacecheck#1{\dimen@=\pagegoal\advance\dimen@ by -\pagetotal
   \ifdim\dimen@<#1 \ifdim\dimen@>0pt \vfil\break \fi\fi}
\def\chapter#1{\par \penalty-300 \vskip\chapterskip
   \spacecheck\chapterminspace
   \chapterreset \titlestyle{\chapterlabel \ #1}
   \nobreak\vskip\headskip \penalty 30000
   \wlog{\string\chapter\ \chapterlabel} }

%
\def\section#1{\par \ifnum\the\lastpenalty=30000\else
   \penalty-200\vskip\sectionskip \spacecheck\sectionminspace\fi
   \wlog{\string\section\ \chapterlabel \the\sectionnumber}
   \global\advance\sectionnumber by 1
   {\ifPhysRev {\centerline {\sectionlabel #1}\par } \else
        { \noindent {\caps \chapterlabel
            \sectionlabel #1}\par } \fi}
   \nobreak\vskip\headskip \penalty 30000 }
\def\ssection#1{\par \ifnum\the\lastpenalty=30000\else
   \penalty-200\vskip\sectionskip \spacecheck\sectionminspace\fi
   \wlog{\string\section\ \chapterlabel \the\sectionnumber}
   \global\advance\sectionnumber by 1  \noindent
   {\S \caps\thinspace\chapterlabel \sectionlabel #1}\par
   \nobreak\vskip\headskip \penalty 30000 }
\def\subsection#1{\par
   \ifnum\the\lastpenalty=30000\else \penalty-100\smallskip \fi
   \noindent\undertext{#1}\enspace \vadjust{\penalty5000}}

\def\undertext#1{\vtop{\hbox{#1}\kern 1pt \hrule}}
\def\APPENDIX#1#2{\par\penalty-300\vskip\chapterskip
   \spacecheck\chapterminspace \chapterreset \xdef\chapterlabel{#1}
   \titlestyle{APPENDIX #2} \nobreak\vskip\headskip \penalty 30000
   \wlog{\string\Appendix\ \chapterlabel} }
\def\Appendix#1{\APPENDIX{#1}{#1}}
\def\appendix{\APPENDIX{A}{}}
%
\def\eqname#1{\relax \ifnum\the\equanumber<0
     \xdef#1{{\noexpand\rm(\number-\equanumber)}}
     \global\advance\equanumber by -1
     \else \global\advance\equanumber by 1
    \xdef#1{{\noexpand\rm(\clabel \number\equanumber)}} \fi}

\def\eqn#1{\eqno\eqname{#1}#1}

\def\eqinsert#1{\noalign{\dimen@=\prevdepth \nointerlineskip
   \setbox0=\hbox to\displaywidth{\hfil #1}
   \vbox to 0pt{\vss\hbox{$\!\box0\!$}\kern-0.5\baselineskip}
   \prevdepth=\dimen@}}
%

%

%

%
%
\def\GENITEM#1;#2{\par \hangafter=0 \hangindent=#1
    \Textindent{#2}\ignorespaces}
\outer\def\newitem#1=#2;{\gdef#1{\GENITEM #2;}}
\newdimen\itemsize		  \itemsize=30pt
\newitem\item=1\itemsize;
\newitem\sitem=1.75\itemsize;	  
\newitem\ssitem=2.5\itemsize;	  
\outer\def\newlist#1=#2&#3&#4;{\toks0={#2}\toks1={#3}%
   \count255=\escapechar \escapechar=-1
   \alloc@0\list\countdef\insc@unt\listcount	 \listcount=0
   \edef#1{\par
      \countdef\listcount=\the\allocationnumber
      \advance\listcount by 1
      \hangafter=0 \hangindent=#4
      \Textindent{\the\toks0{\listcount}\the\toks1}}
   \expandafter\expandafter\expandafter
    \edef\c@t#1{begin}{\par
      \countdef\listcount=\the\allocationnumber \listcount=1
      \hangafter=0 \hangindent=#4
      \Textindent{\the\toks0{\listcount}\the\toks1}}
   \expandafter\expandafter\expandafter
    \edef\c@t#1{con}{\par \hangafter=0 \hangindent=#4 \noindent}
   \escapechar=\count255}
\def\c@t#1#2{\csname\string#1#2\endcsname}
\newlist\point=\Number&.&1.0\itemsize;
\newlist\subpoint=(\alphabetic&)&1.75\itemsize;
\newlist\subsubpoint=(\roman&)&2.5\itemsize;
%

%
%
%
\newcount\referencecount     \referencecount=0
\newif\ifreferenceopen	     \newwrite\referencewrite
\newtoks\rw@toks
\def\NPrefmark#1{\attach{\scriptscriptstyle  #1 ) }}
\let\PRrefmark=\attach
\def\refmark#1{\relax\ifPhysRev\PRrefmark{#1}\else\NPrefmark{#1}\fi}
\def\refend{\refmark{\number\referencecount}}
\newcount\lastrefsbegincount \lastrefsbegincount=0
\def\refsend{\refmark{\count255=\referencecount
   \advance\count255 by-\lastrefsbegincount
   \ifcase\count255 \number\referencecount
   \or \number\lastrefsbegincount,\number\referencecount
   \else \number\lastrefsbegincount-\number\referencecount \fi}}
\def\refch@ck{\chardef\rw@write=\referencewrite
   \ifreferenceopen \else \referenceopentrue
   \immediate\openout\referencewrite=reference.aux \fi}
%
{\catcode`\^^M=\active 
  \gdef\obeyendofline{\catcode`\^^M\active \let^^M\ }}%
%
{\catcode`\^^M=\active 
  \gdef\ignoreendofline{\catcode`\^^M=5}}
{\obeyendofline\gdef\rw@start#1{\def\t@st{#1} \ifx\t@st\blankend%
\endgroup \@sf \relax \else \ifx\t@st\bl@nkend \endgroup \@sf \relax%
\else \rw@begin#1
\backtotext
\fi \fi } }
{\obeyendofline\gdef\rw@begin#1
{\def\n@xt{#1}\rw@toks={#1}\relax%
\rw@next}}
\def\blankend{}
{\obeylines\gdef\bl@nkend{
}}
\newif\iffirstrefline  \firstreflinetrue
\def\rwr@teswitch{\ifx\n@xt\blankend \let\n@xt=\rw@begin %
 \else\iffirstrefline \global\firstreflinefalse%
\immediate\write\rw@write{\noexpand\obeyendofline \the\rw@toks}%
\let\n@xt=\rw@begin%
      \else\ifx\n@xt\rw@@d \def\n@xt{\immediate\write\rw@write{%
	\noexpand\ignoreendofline}\endgroup \@sf}%
	     \else \immediate\write\rw@write{\the\rw@toks}%
	     \let\n@xt=\rw@begin\fi\fi \fi}
\def\rw@next{\rwr@teswitch\n@xt}
\def\rw@@d{\backtotext} \let\rw@end=\relax
\let\backtotext=\relax

\newdimen\refindent	\refindent=30pt
\def\refitem#1{\par \hangafter=0 \hangindent=\refindent \Textindent{#1}}
\def\REFNUM#1{\space@ver{}\refch@ck \firstreflinetrue%
 \global\advance\referencecount by 1 \xdef#1{\the\referencecount}}
\def\refnum#1{\space@ver{}\refch@ck \firstreflinetrue%
 \global\advance\referencecount by 1 \xdef#1{\the\referencecount}\refend}

\def\REF#1{\REFNUM#1%
 \immediate\write\referencewrite{%
 \noexpand\refitem{#1.}}%
\begingroup\obeyendofline\rw@start}
\def\ref{\refnum\?%
 \immediate\write\referencewrite{\noexpand\refitem{\?.}}%
\begingroup\obeyendofline\rw@start}
\def\Ref#1{\refnum#1%
 \immediate\write\referencewrite{\noexpand\refitem{#1.}}%
\begingroup\obeyendofline\rw@start}
\def\REFS#1{\REFNUM#1\global\lastrefsbegincount=\referencecount
\immediate\write\referencewrite{\noexpand\refitem{#1.}}%
\begingroup\obeyendofline\rw@start}
\newcount\figurecount	  \figurecount=0
\newif\iffigureopen	  \newwrite\figurewrite
\def\figch@ck{\chardef\rw@write=\figurewrite \iffigureopen\else
   \immediate\openout\figurewrite=figures.aux
   \figureopentrue\fi}
\def\FIGNUM#1{\space@ver{}\figch@ck \firstreflinetrue%
 \global\advance\figurecount by 1 \xdef#1{\the\figurecount}}
\def\FIG#1{\FIGNUM#1
   \immediate\write\figurewrite{\noexpand\refitem{#1.}}%
   \begingroup\obeyendofline\rw@start}
\def\fig{\FIGNUM\? fig.~\?%
\immediate\write\figurewrite{\noexpand\refitem{\?.}}%
\begingroup\obeyendofline\rw@start}
\def\figure{\FIGNUM\? figure~\?
   \immediate\write\figurewrite{\noexpand\refitem{\?.}}%
   \begingroup\obeyendofline\rw@start}
\def\Fig{\FIGNUM\? Fig.~\?%
\immediate\write\figurewrite{\noexpand\refitem{\?.}}%
\begingroup\obeyendofline\rw@start}
\def\Figure{\FIGNUM\? Figure~\?%
\immediate\write\figurewrite{\noexpand\refitem{\?.}}%
\begingroup\obeyendofline\rw@start}
\newcount\tablecount	 \tablecount=0
\newif\iftableopen	 \newwrite\tablewrite
\def\tabch@ck{\chardef\rw@write=\tablewrite \iftableopen\else
   \immediate\openout\tablewrite=tables.aux
   \tableopentrue\fi}
\def\TABNUM#1{\space@ver{}\tabch@ck \firstreflinetrue%
 \global\advance\tablecount by 1 \xdef#1{\the\tablecount}}
\def\TABLE#1{\TABNUM#1
   \immediate\write\tablewrite{\noexpand\refitem{#1.}}%
   \begingroup\obeyendofline\rw@start}
\def\Table{\TABNUM\? Table~\?%
\immediate\write\tablewrite{\noexpand\refitem{\?.}}%
\begingroup\obeyendofline\rw@start}
%

%
%
%
\def\masterreset{\global\pagenumber=1 \global\chapternumber=0
   \global\equanumber=0 \global\sectionnumber=0
   \global\referencecount=0 \global\figurecount=0 \global\tablecount=0 }
\def\FRONTPAGE{\ifvoid255\else\vfill\penalty-2000\fi
      \masterreset\global\frontpagetrue
      \global\lastf@@t=0 \global\footsymbolcount=0}

\def\paperstyle{\letterstylefalse\normalspace\papersize}
\def\letterstyle{\letterstyletrue\singlespace\lettersize}
%
%
\def\papersize{\hsize=158mm\vsize=226mm\hoffset=0mm\voffset=6mm
  	       \skip\footins=\bigskipamount}
%
\def\lettersize{\hsize=35.0pc\vsize=50.0pc\hoffset=1.2pc\voffset=2.8pc
   \skip\footins=\smallskipamount \multiply\skip\footins by 3 }
%
\paperstyle   
%
%
\def\MEMO{\letterstyle\FRONTPAGE \letterfrontheadline={\hfil}
    \line{\quad\fourteenrm KEK MEMORANDUM\hfil\twelverm\the\date\quad}
    \medskip \memod@f}

\def\memit@m#1{\smallskip \hangafter=0 \hangindent=1in
      \Textindent{\caps #1}}
\def\memod@f{\xdef\to{\memit@m{To:}}\xdef\from{\memit@m{From:}}%
     \xdef\topic{\memit@m{Topic:}}\xdef\subject{\memit@m{Subject:}}%
     \xdef\rule{\bigskip\hrule height 1pt\bigskip}}
\memod@f
%

%
\newskip\lettertopfil
\lettertopfil = 0pt plus 1.5in minus 0pt
\newskip\letterbottomfil
\letterbottomfil = 0pt plus 2.3in minus 0pt
\newskip\spskip \setbox0\hbox{\ } \spskip=-1\wd0
\def\addressee#1{\medskip\rightline{\the\date\hskip 30pt} \bigskip
   \vskip\lettertopfil
   \ialign to\hsize{\strut ##\hfil\tabskip 0pt plus \hsize \cr #1\crcr}
   \medskip\noindent\hskip\spskip}
\newskip\signatureskip	     \signatureskip=40pt
\def\signed#1{\par \penalty 9000 \bigskip \dt@pfalse
  \everycr={\noalign{\ifdt@p\vskip\signatureskip\global\dt@pfalse\fi}}
  \setbox0=\vbox{\singlespace \halign{\tabskip 0pt \strut ##\hfil\cr
   \noalign{\global\dt@ptrue}#1\crcr}}
  \line{\hskip 2cm\hskip 0.5\hsize minus 0.5\hsize \box0\hfil\hss} \medskip}

\def\endletter{\ifnum\pagenumber=1 \vskip\letterbottomfil\supereject
\else \vfil\supereject \fi}
\newbox\letterb@x
\def\lettertext{\par\unvcopy\letterb@x\par}
\def\multiletter{\setbox\letterb@x=\vbox\bgroup
      \everypar{\vrule height 1\baselineskip depth 0pt width 0pt }
      \singlespace \topskip=\baselineskip }
\def\letterend{\par\egroup}
%
%
%
\newskip\frontpageskip
\newtoks\pubtype
\newtoks\Pubnum
\newtoks\pubnum
\newif\ifp@bblock  \p@bblocktrue
\def\PH@SR@V{\doubl@true \baselineskip=18.0pt plus 0.2pt minus 0.1pt
	     \parskip= 3pt plus 2pt minus 1pt }
\def\PHYSREV{\paperstyle\PhysRevtrue\PH@SR@V}
\def\titlepage{\FRONTPAGE\paperstyle\ifPhysRev\PH@SR@V\fi
          \ifp@bblock\p@bblock\fi}
\def\nopubblock{\p@bblockfalse}
\def\endpage{\vfil\break}
\frontpageskip=1\medskipamount plus .5fil
\pubtype={ }
\newtoks\publevel
\publevel={Report}   
\Pubnum={\the\pubnum}
\pubnum={0000}
\def\p@bblock{\begingroup \tabskip=\hsize minus \hsize
   \baselineskip=1.5\ht\strutbox \topspace-2\baselineskip
   \halign to\hsize{\strut ##\hfil\tabskip=0pt\crcr
   \the\Pubnum\cr \the\date\cr }\endgroup}
%
\def\title#1{\vskip\frontpageskip\vfill
   {\fourteenbf\titlestyle{#1}}\vskip\headskip\vfill }
\def\author#1{\vskip\frontpageskip\titlestyle{\twelvecp #1}\nobreak}

%
\def\address#1{\par\kern 5pt \titlestyle{\twelvepoint\sl #1}}
\def\andaddress{\par\kern 5pt \centerline{\sl and} \address}

%
\def\abstract#1{\vfill\vskip\frontpageskip
    \hbox{{\fourteencp Abstract}\hfill}\vskip\headskip#1\vfill\endpage}

%

%
%
%

\def\\{\relax\ifmmode\backslash\else$\backslash$\fi}
\def\globaleqnumbers{\relax\if\equanumber<0\else\global\equanumber=-1\fi}
\def\nextline{\unskip\nobreak\hskip\parfillskip\break}

\def\journal#1&#2 #3 (#4){\rm #1  \unskip {\bf #2}\rm, #3~(19#4)}

\def\topspace{\hrule height 0pt depth 0pt \vskip}

\def\bra#1{\left\langle #1\right\vert}
\def\ket#1{\left\vert #1\right\rangle}

\let\int=\intop		
\def\prop{\mathrel{{\mathchoice{\pr@p\scriptstyle}{\pr@p\scriptstyle}{
		\pr@p\scriptscriptstyle}{\pr@p\scriptscriptstyle} }}}
\def\pr@p#1{\setbox0=\hbox{$\cal #1 \char'103$}
   \hbox{$\cal #1 \char'117$\kern-.4\wd0\box0}}
\def\lsim{\mathrel{\mathpalette\@versim<}}
\def\gsim{\mathrel{\mathpalette\@versim>}}
\def\@versim#1#2{\lower0.2ex\vbox{\baselineskip\z@skip\lineskip\z@skip
  \lineskiplimit\z@\ialign{$\m@th#1\hfil##\hfil$\crcr#2\crcr\sim\crcr}}}
%
%
%
\let\sec@nt=\sec
\def\sec{\relax\ifmmode\let\n@xt=\sec@nt\else\let\n@xt\section\fi\n@xt}
\def\obsolete#1{\message{Macro \string #1 is obsolete.}}
\def\firstsec#1{\obsolete\firstsec \section{#1}}
\def\firstsubsec#1{\obsolete\firstsubsec \subsection{#1}}
\def\thispage#1{\obsolete\thispage \global\pagenumber=#1\frontpagefalse}
\def\thischapter#1{\obsolete\thischapter \global\chapternumber=#1}
\def\nextequation#1{\obsolete\nextequation \global\equanumber=#1
   \ifnum\the\equanumber>0 \global\advance\equanumber by 1 \fi}
\def\BOXITEM{\afterassigment\B@XITEM\setbox0=}
\def\B@XITEM{\par\hangindent\wd0 \noindent\box0 }
%

%
\catcode`@=12 
%


\relax

\def\yen{\hbox{Y\kern-0.75em =}}
%
%
\catcode`@=11
%
%
\font\fourteenmib=cmmib10 scaled\magstep2    \skewchar\fourteenmib='177
\font\twelvemib=cmmib10 scaled\magstep1	    \skewchar\twelvemib='177
\font\elevenmib=cmmib10 scaled\magstephalf   \skewchar\elevenmib='177
\font\tenmib=cmmib10			    \skewchar\tenmib='177
%
\font\fourteenbsy=cmbsy10 scaled\magstep2     \skewchar\fourteenbsy='60
\font\twelvebsy=cmbsy10 scaled\magstep1	      \skewchar\twelvebsy='60
\font\elevenbsy=cmbsy10 scaled\magstephalf    \skewchar\elevenbsy='60
\font\tenbsy=cmbsy10			      \skewchar\tenbsy='60
%
\newfam\mibfam
\def\samef@nt{\relax \ifcase\f@ntkey \rm \or\oldstyle \or\or
	 \or\it \or\sl \or\bf \or\tt \or\caps \or\mib \fi }
\def\fourteenpoint{\relax
    \textfont0=\fourteenrm	    \scriptfont0=\tenrm
    \scriptscriptfont0=\sevenrm
     \def\rm{\fam0 \fourteenrm \f@ntkey=0 }\relax
    \textfont1=\fourteeni	    \scriptfont1=\teni
    \scriptscriptfont1=\seveni
     \def\oldstyle{\fam1 \fourteeni\f@ntkey=1 }\relax
    \textfont2=\fourteensy	    \scriptfont2=\tensy
    \scriptscriptfont2=\sevensy
    \textfont3=\fourteenex     \scriptfont3=\fourteenex
    \scriptscriptfont3=\fourteenex
    \def\it{\fam\itfam \fourteenit\f@ntkey=4 }\textfont\itfam=\fourteenit
    \def\sl{\fam\slfam \fourteensl\f@ntkey=5 }\textfont\slfam=\fourteensl
    \scriptfont\slfam=\tensl
    \def\bf{\fam\bffam \fourteenbf\f@ntkey=6 }\textfont\bffam=\fourteenbf
    \scriptfont\bffam=\tenbf	 \scriptscriptfont\bffam=\sevenbf
    \def\tt{\fam\ttfam \twelvett \f@ntkey=7 }\textfont\ttfam=\twelvett
    \h@big=11.9\p@{} \h@Big=16.1\p@{} \h@bigg=20.3\p@{} \h@Bigg=24.5\p@{}
    \def\caps{\fam\cpfam \twelvecp \f@ntkey=8 }\textfont\cpfam=\twelvecp
    \setbox\strutbox=\hbox{\vrule height 12pt depth 5pt width\z@}
    \def\mib{\fam\mibfam \fourteenmib \f@ntkey=9 }
    \textfont\mibfam=\fourteenmib      \scriptfont\mibfam=\tenmib
    \scriptscriptfont\mibfam=\tenmib
    \samef@nt}
\def\twelvepoint{\relax
    \textfont0=\twelverm	  \scriptfont0=\ninerm
    \scriptscriptfont0=\sixrm
     \def\rm{\fam0 \twelverm \f@ntkey=0 }\relax
    \textfont1=\twelvei		  \scriptfont1=\ninei
    \scriptscriptfont1=\sixi
     \def\oldstyle{\fam1 \twelvei\f@ntkey=1 }\relax
    \textfont2=\twelvesy	  \scriptfont2=\ninesy
    \scriptscriptfont2=\sixsy
    \textfont3=\twelveex	  \scriptfont3=\twelveex
    \scriptscriptfont3=\twelveex
    \def\it{\fam\itfam \twelveit \f@ntkey=4 }\textfont\itfam=\twelveit
    \def\sl{\fam\slfam \twelvesl \f@ntkey=5 }\textfont\slfam=\twelvesl
    \scriptfont\slfam=\ninesl
    \def\bf{\fam\bffam \twelvebf \f@ntkey=6 }\textfont\bffam=\twelvebf
    \scriptfont\bffam=\ninebf	  \scriptscriptfont\bffam=\sixbf
    \def\tt{\fam\ttfam \twelvett \f@ntkey=7 }\textfont\ttfam=\twelvett
    \h@big=10.2\p@{}
    \h@Big=13.8\p@{}
    \h@bigg=17.4\p@{}
    \h@Bigg=21.0\p@{}
    \def\caps{\fam\cpfam \twelvecp \f@ntkey=8 }\textfont\cpfam=\twelvecp
    \setbox\strutbox=\hbox{\vrule height 10pt depth 4pt width\z@}
    \def\mib{\fam\mibfam \twelvemib \f@ntkey=9 }
    \textfont\mibfam=\twelvemib	    \scriptfont\mibfam=\tenmib
    \scriptscriptfont\mibfam=\tenmib
    \samef@nt}
\def\tenpoint{\relax
    \textfont0=\tenrm	       \scriptfont0=\sevenrm
    \scriptscriptfont0=\fiverm
    \def\rm{\fam0 \tenrm \f@ntkey=0 }\relax
    \textfont1=\teni	       \scriptfont1=\seveni
    \scriptscriptfont1=\fivei
    \def\oldstyle{\fam1 \teni \f@ntkey=1 }\relax
    \textfont2=\tensy	       \scriptfont2=\sevensy
    \scriptscriptfont2=\fivesy
    \textfont3=\tenex	       \scriptfont3=\tenex
    \scriptscriptfont3=\tenex
    \def\it{\fam\itfam \tenit \f@ntkey=4 }\textfont\itfam=\tenit
    \def\sl{\fam\slfam \tensl \f@ntkey=5 }\textfont\slfam=\tensl
    \def\bf{\fam\bffam \tenbf \f@ntkey=6 }\textfont\bffam=\tenbf
    \scriptfont\bffam=\sevenbf	   \scriptscriptfont\bffam=\fivebf
    \def\tt{\fam\ttfam \tentt \f@ntkey=7 }\textfont\ttfam=\tentt
    \def\caps{\fam\cpfam \tencp \f@ntkey=8 }\textfont\cpfam=\tencp
    \setbox\strutbox=\hbox{\vrule height 8.5pt depth 3.5pt width\z@}
    \def\mib{\fam\mibfam \tenmib \f@ntkey=9 }
    \textfont\mibfam=\tenmib   \scriptfont\mibfam=\tenmib
    \scriptscriptfont\mibfam=\tenmib
    \samef@nt}
%
%
\Twelvepoint
\catcode`@=12
%
%
\everyjob{\input mydef
 \message{PHYZZX macro created by V. K.;modified by H.W., K-I. A. & Y. T.
June '88}}
\message{created by V. K. modified by H.W., K-I. A. & Y. T.}
%
\def\del{\partial}
\def\half{{1 \over 2}}
\def\Lag{{\cal L}}              
\def\Ham{{\cal H}}              
%
\def\bra#1{\langle #1 |}
\def\ket#1{| #1 \rangle}

\def\dslash{\del\kern-0.55em\raise 0.14ex\hbox{/}}
\def\Aslash{A\kern-0.6em\raise 0.14ex\hbox{/}}
%
%
%
%
%

\def\gtsim{\mathrel{\hbox{\raise0.2ex
\hbox{$>$}\kern-0.75em\raise-0.9ex\hbox{$\sim$}}}}
\def\ltsim{\mathrel{\hbox{\raise0.2ex
\hbox{$<$}\kern-0.75em\raise-0.9ex\hbox{$\sim$}}}}
%
%
\def\tdots{ \hskip 0.0pt \raise0.1ex\hbox{.} \kern -0.24em \raise
1.1ex\hbox{.} \kern -0.25em \raise 0.1ex\hbox{.} \   }
\def\trdots{ \hskip 0.0pt \raise 1.1ex\hbox{.} \kern -0.24em \raise
0.1ex\hbox{.} \kern -0.25em \raise 1.1ex\hbox{.} \  }
%
\def\therefore{ \hskip 0.0pt \raise0.1ex\hbox{.} \mkern -0.24mu \raise
1.1ex\hbox{.} \mkern -0.25mu \raise 0.1ex\hbox{.} \   }
\def\because{ \hskip 0.0pt \raise 1.1ex\hbox{.} \mkern -0.24mu \raise
0.1ex\hbox{.} \mkern -0.25mu \raise 1.1ex\hbox{.} \  }
%
%
%
\message{created by K.Funakubo, updated on July, 1990}
\input tables
%
%
\def\gtsim{\mathrel{\hbox{\raise0.2ex
\hbox{$>$}\kern-0.75em\raise-0.9ex\hbox{$\sim$}}}}
\def\ltsim{\mathrel{\hbox{\raise0.2ex
\hbox{$<$}\kern-0.75em\raise-0.9ex\hbox{$\sim$}}}}
\def\inprod#1#2{\langle #1 | #2 \rangle}

\def\k{\kappa}

\def\sn{{\rm sn}}            
%
%
\baselineskip 15.0pt
\pubnum={  }
\date={  }
\hfill{\vbox{\hbox{SAGA--HE--46}
             \hbox{KYUSHU--HET--6}
             \hbox{October, 1992}}}
\baselineskip=24.2pt plus 0.2pt minus 0.1pt
\titlepage
\font\tt=cmbx10 scaled\magstep3

\vskip 2.5cm
\centerline{\tt Scattering with Baryon Number Violation}\par
\vskip 5mm
\centerline{\tt --- The Case of Higgs Particle Production ---}
\vskip 1.0cm
\centerline {Koichi FUNAKUBO, Shoichiro OTSUKI$^\dagger$}
\centerline {Kazunori TAKENAGA$^\dagger$ and Fumihiko TOYODA$^\ddagger$}
\vskip 5mm
\centerline {\sl Department of Physics, Saga University,}
\centerline {\sl Saga 840, Japan}
\centerline {\sl $^\dagger$Department of Physics, Kyushu University 33,}
\centerline {\sl Fukuoka 812, Japan}
\centerline {\sl $^\ddagger$Department of Liberal Arts, Kinki University
 in Kyushu,}
\centerline {\sl Iizuka 820, Japan}
%
\abstract{\noindent
A formalism based on path-integral expression of time-evolution operator
during tunneling at a finite energy proposed by the authors is applied to
$SU(2)$
gauge-Higgs system to produce Higgs particles with $\Delta B=1$.
Instead of starting from instanton tunneling at the zero energy, a classical
bounce
solution
giving sphaleron (instanton) action at high (low) energies is used as the
tunneling configuration.
Fourier transform of the bounce configuration in coherent state expression
at the entrance and exit of the
tunneling plays an important role. Numerical results at various energies
for $M_H/M_W=1 \sim 2 $ are given.
Though the cross section with $\Delta B=1$ results from a severe
cancellation of several large quantities in the
leading order as occured in the instanton calculus, it seems unlikely that
the cross section grows as largely as to reach unitarity bound at energies
$E \leq E_{sph}$. It is pointed out that the actual value $g^2=0.418$ of the
$SU(2)$
gauge coupling constant may be too large to take the weak coupling limit.}
%
%
%
%
\REF\tHooft{G.'t Hooft, Phys.Rev.Lett.{\bf 37}(1976),8;
                        Phys.Rev.{\bf D14}(1976),3432;\nextline
                        (E)Phys.Rev.{\bf D18}(1978),2199.}
\REF\Manton{N.S.Manton, Phys.Rev.{\bf D28}(1983),2019.\nextline
            F.R.Klinkhamer and N.S.Manton, Phys.Rev.{\bf D30}(1984),2212.}
\REF\krs{V.A.Kuzumin, V.A.Rubakov and M.E.Shaposhnikov,
  Phys.Lett.{\bf B155}(1985),36.}
\REF\Ring{A.Ringwald, Nucl.Phys.{\bf B330}(1990),1.}
\REF\Espi{O.Espinpsa, Nucl.Phys.{\bf B343}(1990),310.}
\REF\ag{H.Aoyama and H.Goldberg, Phys.Lett.{\bf B188}(1987),506.}
\REF\Mattis{See, for example, M.P.Mattis,LA-UR-91-2926(1991).}
\REF\fott{K.Funakubo, S.Otsuki, K.Takenaga and F.Toyoda, Prog.Theor.Phys.
{\bf 87}\nextline
(1992),663. This paper is referred to as I.}
\REF\Coleman{C.G,Callan, Jr. and S.Coleman, Phys.Rev.{\bf D16}(1977),1762.}
\REF\BanksBenderWu{T.Banks, C.M.Bender and T.T.Wu,
                   Phys.Rev.{\bf D8}(1973),3346.}
\REF\GervaisSakita{J.-L.Gervais and B.Sakita, Phys.Rev.{\bf D16}(1977),3507.}
\REF\agr{H.Aoyama, H.Goldberg and Z.Ryzak, Phys.Rev.Lett,{\bf
60}(1988),1902.}
\REF\Klauder{J.R.Klauder, Phys.Rev.Lett.{\bf 56}(1986),897;\nextline
             A.Voros, Phys.Rev.{\bf A40}(1989),6814.}
\REF\shap{M.E.Shaposhnikov, Phys.Lett.{\bf B242}(1990),493.}
\REF\bfds{T.Banks, G.Farrar, M.Dine, D.Karabali and B.Sakita, Nucl.Phys.{\bf
B347}
(1990),581.}
\REF\KRT{S.Yu.Klhebnikov, V.A.Rubakov and P.G.Tinyakov, Nucl.Phys.{\bf
B367}(1991),334
.}
\REF\cah{K.Cahill, Phys.Lett.{\bf B53}(1974),174.}
\REF\McLaughlin{D.W.McLaughlin, J.Math.Phys.{\bf 13}(1972),1099.}
\REF\BRS{T.Kugo and I.Ojima, Prog.Theor.Phys.Suppl.{\bf 66}(1979), 1.}
\REF\ry{B.Ratra and L.G.Yaffe, Phys.Lett.{\bf B205}(1988),57.}
\REF\fot{K.Funakubo, S.Otsuki and F.Toyoda, Prog.Theor.Phys.{\bf 84},1196
(1990).}
\REF\Yaffe{L.G.Yaffe, Phys.Rev.{\bf D40}(1989),3463.}
\REF\zad{J.Zadrozny, Phys.Lett.{\bf B284}(1992),88.}
%

%
%
%
\chapter{Introduction}\par
It is well-known that chiral anomaly in baryon number current of electroweak
theory violates baryon number ($\Delta B\not =0$) given by topologically
nontrivial background gauge fields through quantum
tunneling. \refmark{\tHooft} At low energies or low temperatures
the probability of
$\Delta B\not =0$ processes through instanton tunneling is known to be
unobservably
small.
Sphaleron\refmark{\Manton} is a saddle-point solution to equations of motion
in the
electroweak theory, and its mass, $E_{sph} \sim M_W/\alpha_{W}$, gives
the barrier hight between the topologically inequivalent vacua.
Though there remains problem of non-equilibrium process, the transition
rate with $\Delta B\not=0$ at high temperatures $\sim E_{sph}$ is commonly
believed
to be significant\refmark{\krs} and to be crucial for baryon asymmetry of the
universe. \par
On the other hand, after works by Ringwald\refmark{\Ring} and
Espinosa,\refmark
{\Espi} which succeeded a pioneering one by Aoyama and Goldberg,\refmark{\ag}
that the $\Delta B\not=0$ process
might be detected in future colliders, the scattering cross section at high
energies
has been
evaluated by many authors. Almost all of these analyses are based on
instanton
calculus and the LSZ formalism, by which the cross section in the leading
order
consists of three factors; the instanton suppression
factor, $({\rm residue})^{n+2}$ and the $n$-body phase space volume with $n$
denoting the
number of produced particles. The residue here is meant by
that of the euclidean propagator in the instanton background of the particle.
For $n$ large enough, the last two factors overcome the
first one and raise the cross
section even to the unitarity bound. It may be natural then that possible
corrections
to the leading order are
large and tend to hinder the rise of the cross section. We are in a
frustrated
situation, as the corrections may be comparable with the leading order
contribution
so that no reliable answer has been obtained.\refmark{\Mattis} \par
In a previous paper,\refmark{\fott} the authors have proposed an alternative
approach to treat the $\Delta B\not=0$ high energy scattering, by noticing
that
{\it the process is basically a tunneling process traversing
not far from the barrier top so that starting from the instanton tunneling
may not be appropriate}. We have applied the
formalism to a simple model accompanying both the instanton and the
sphaleron, $O(3)$
nonlinear sigma model in $1+1$ dimensions.
In this paper we apply our formalism to more realistic model, $3+1$
dimensional
$SU(2)$ gauge-Higgs system, clarify structure of the amplitude, and
numerically analyze the production cross
section of $n$ Higgs particles from the two ones. \par
For tunneling configuration at non-zero energies, we use a classical bounce
solution
\refmark{\Coleman}
instead of the instanton, because the tunneling exponent $W(E)$ at $E \not=0$
is not given by the instanton
action but rather by the Legendre transform of the bounce action
as is obvious in quantum mechanics of one degree of freedom. Here the
tunneling exponent is given by $ W(E)
\propto\int_{x_1(E)}^{x_2(E)}dx\sqrt{V(x)-E}$
where
$x_{1,2}$ are the turning points at energy $E$. At $E=0$, $W(0)$ coincides
with
the instanton action while at $E\not=0$, the transform leads to
$W(E)=S_b(T)+T(E)E$
where $T(E)$ is the half period of the bounce motion and $S_b$ is the half of
the
bounce action.
Within the WKB approximation, the
tunneling process in many-dimensional quantum mechanics and field theory may
be forced
to reduce to an
essentially one-dimensional problem around some classical
configurations.\refmark{\BanksBenderWu,\GervaisSakita}
Once we find a classical
bounce solution employing some ansatz,
the path-integral expression of time-evolution operator during tunneling
at a given energy would be dominated by the configuration. \par
For the $\Delta B=1$ process, the non-contractible loop parameter $\mu$
, that connects a neighboring pair of topologically inequivalent vacua,
\refmark{\Manton} may be the most appropriate dynamical
variable describing the tunneling through the sphaleron barrier.
A variational parameter is introduced to compensate ignorance of other
degrees of
dynamical freedom in the tunneling (deformed sphaleron).
Actually the classical bounce solution of this type of reduced
model\refmark{\agr} does bridge the gap between instanton and the sphaleron
smoothly.
Coherent state expression of the entrance and exit of the tunneling
mediated by the bounce solution helps us to avoid singularities at the
turning points,\refmark{\Klauder} and plays an important role instead of the
residue in the instanton background. \par
There are some works of
interest by Shaposhnikov,\refmark{\shap} Banks {\it et al}.\refmark{\bfds}
and
Khlebnikov. {\it et al}\refmark{\KRT}, which would be
close in spirit to ours in the meaning that they do not start from the zero
energy
instanton but treat the tunneling at a finite energy from the first. We will
make
some comments on them in due places. \par
This paper is organized as follows.
In section 2 we briefly review our formalism.
In section 3 we show
calculations by the reduced model and clarify
structure of the scattering amplitude.
In section 4 numerical
analyses of the cross section with $\Delta B =1$ at various energies
and for mass ratio $M_H/M_W=1 \sim 2$ are presented.
The high energy cross section  results from a severe
cancellation of several large quantities in the leading order
as occured in the instanton calculus. If we take the face values, it seems
unlikely that the cross
section grows as largely as to reach the unitarity bound at energies $E\leq
E_{sph}$.
Section 5 is devoted to concluding remarks. The actual value $g^2=0.418$ of
the
$SU(2)$ gauge coupling constant may be too large to make the weak coupling
approximation. Gauge fixing conditions in $R_{\xi}$
gauge of the reduced model are summarized in Appendix.
\chapter{Scattering amplitude through bounce configuration}
\par
Characteristic points of our formalism are as follows.
\section{$S$-matrix element}\noindent
The $S$-operator is defined by
$$
     \hat S = \lim_{t_F\rightarrow \infty \atop t_I\rightarrow -\infty}
														{\rm e}^{i\hat H_0t_F/\hbar}\>{\rm e}^{-i\hat H(t_F-t_I)/\hbar}
              \>{\rm e}^{-i\hat H_0t_I/\hbar},
\eqn\Ia$$
where $\hat H$ is the hamiltonian and $\hat H_0$ is the free part of it.
For simplicity we ignore initial- and final-state interactions. Then we have
$$	  {\rm e}^{-i\hat H(t_F-t_I)/\hbar}
  \simeq {\rm e}^{-i\hat H_0(t_F-X^0-T/2)/\hbar}\>{\rm e}^{-i\hat H T/\hbar}
         \>{\rm e}^{-i\hat H_0(X^0-T/2-t_I)/\hbar},
\eqn\Ib$$
where $T(E)$ is the half period of the bounce motion and $X^0$ is
the time-like
center of it. \par
%
%
By inserting the identity operator in
terms of coherent state, the above $S$-operator is written as
$$\eqalign{
  \hat S \simeq\; & {\rm e}^{i\hat H_0(X^0+T/2)/\hbar}\int
            \prod_{t=X^0-T/2}^{X^0+T/2}\Biggl[{d\Phi(t)d\Pi(t)\over2\pi\hbar}
             \Biggr] \ket{\Phi(X^0+T/2),\Pi(X^0+T/2)}           \cr
           & \times \exp\Bigl\{{i\over\hbar}\int_{t=X^0-T/2}^{X^0+T/2} d^d
x\,
             [ \Pi(x)\dot\Phi(x) - \Ham( \Phi,\Pi ) ] \Bigr\}           \cr
           & \times \bra{\Phi(X^0-T/2),\Pi(X^0-T/2)}
             \>{\rm e}^{-i\hat H_0(X^0-T/2)/\hbar},               \cr
           }\eqn\Ic$$
where the state $\ket{\phi(x),\pi(x)}$ is defined by
$$
  \ket{\phi(x),\pi(x)}=\exp\Bigl\{-{i\over\hbar}\int d^{d-1}{\bf x} \bigl[
                   \phi(x)\hat\Pi(0,{\bf x})-\pi(x)\hat\Phi(0,{\bf x})
                       \bigr] \Bigr\}\ket{0},
\eqn\Id$$
with $\ket0$ being vacuum annihilated by any annihilation operator $\hat
a_i({\bf k})$
which composes $\hat\Phi(0,{\bf x})$.
Here $\hat\Phi$ and $\hat\Pi$ stand for the canonical variables in the
theory.
The reasons we use coherent state here are not only that the state is the one
with minimal quantum uncertainty so that it will be suited to describe
classical configurations,\refmark{\cah} but also that the WKB approximation
in terms
of it is
global and uniform in contrast to the usual one in coordinate representation
which
brings singularities into wave functions at the turning
points.\refmark{\Klauder}
Since
the residual states
in \Ic\ are just those on the turning points, the coherent state
representation will be very suitable. \par
By extracting collective coordinates (such as $X^0$ and those for internal
symmetry), we recast the expression \Ic\ into an appropriate form suited for
expansion about
the classical solution. The $S$-matrix element is given by taking matrix
element
of the $S$-operator between the initial- and final-asymptotic states which
are the
Fock states with some definite baryon number {\it B} such as
${\ket{\rm asym}}_B=\hat a^\dagger({\bf k}_1) \hat a^\dagger
({\bf k}_2)\cdots$ $\cdots \hat a^\dagger({\bf k}_n)\ket{0}_B$ where the
creation operators are the same as those constructing the coherent state
within the
approximation
that the initial- final-state interactions are ignored. Then the $S$-matrix
element
in the leading order of the WKB approximation ($\hbar \sim 0$)  reads
\foot{For more detailed, see I.}
$$\eqalign{
 S_{fi}
\simeq& \>(2\pi\hbar)^d \delta^d( P_f-P_i)\,{\rm e}^{ - W(E)/\hbar}
                                                  \cr
      & \times \inprod{f}{\phi_c(T/2,{\bf x}), \pi_c(T/2,{\bf x}) }_{B=1}
            \inprod{ \phi_c(-T/2,{\bf x}), \pi_c(-T/2,{\bf x})}{i}_{B=0}. \cr
}\eqn\Ie$$
The tunneling exponent $W(E)$ is obtained by
continuing the time variable to negative imaginary axis in path integral
in \Ic ;
\refmark{\McLaughlin}
$$
     {i\over\hbar}\bigl\{ S[\phi_c,\pi_c]+ET(E) \bigr\} \rightarrow
     -{1\over\hbar} W(E).
    \eqn\If$$
For a bounce solution,
$\phi(\pm T/2)$ is the edge of it and $\pi(\pm T/2)=0$ is sitting at the
turning
points.
The energy-momentum conserving $\delta$ function in the above expression
comes
from integrating out
the collective coordinates of translation. In  $SU(2)$ gauge-Higgs
system in the $R_{\xi}$ gauge, the internal symmetry comes from global
$SU(2)$
transformation; $A^a_{\mu}$  and BRS\refmark{\BRS} quartet
($ \chi^a, C^a, \bar C^a, B^a $) transforming
as $SU(2)$ triplets. If we set up
the initial and final
states to be eigenstates of conserved charges which generate the
corresponding
symmetries, integration of the collective coordinates yields $\delta$
functions of
conservation of the symmetries.
The inner products between the Fock states and the coherent states are given
by the following formula;
$$
   \bra{0} \> \hat a({\bf k}_1)\hat a({\bf k}_2)\cdots
    \hat a({\bf k}_n) \ket{\phi(x),\pi(x)}
 = {\rm e}^{ -\half\int d{\bf k} |\alpha({\bf k})|^2 }
    \alpha({\bf k}_1)\alpha({\bf k}_2)\cdots
    \alpha({\bf k}_n),
\eqn\Ig$$
where $\alpha({\bf k})$ is defined by
$$
 \alpha({\bf k}) =
   \int{ {d^{d-1}{\bf x}} \over \sqrt{(2\pi)^{d-1}2\hbar\omega_{\bf k}} }
   \Bigl[ \omega_{\bf k}\phi(x) + i \pi(x)\Bigr]\,
   {\rm e}^{-i{\bf k}\cdot{\bf x}}
\eqn\Ih$$
with $\omega_{\bf k}$ being energy of asymptotic particle with momentum ${\bf
k}$.
Hence one can evaluate the $S$-matrix element once one has the Fourier
transform of
the classical configuration.
\section{Reduced model}\noindent
In general, a classical solution is given in some gauge and we perform
expansion
around it. So one must care about gauge-independence of $S$-matrix.
The BRS formalism\refmark{\BRS} is one of systematic methods which
automatically
gives gauge-independent $S$-matrix elements. We choose the $R_{\xi}$ gauge,
which
is suited to a problem with a classical background. \par
In this gauge the lagrangian
is given by
$${
\Lag={-1\over4}F^{a}_{\mu \nu}F^{a\mu \nu}+{\vert D_{\mu}\Phi \vert}^2 -
{\lambda \over 2}\bigl({\vert \Phi\vert}^2-{v^2 \over 2 }\bigr)^2 +
\Lag_{R_{\xi}}}.
\eqn\Ii$$
Here the field strength and covariant derivative are defined respectively as
$$ \eqalign{ F^{a}_{\mu \nu}&\equiv\del_\mu A^{a}_{\nu} - \del_\nu
A^{a}_{\mu}
+g\epsilon_{abc}A^{b}_{\mu} A^{c}_{\nu},\cr
D_\mu \Phi     &\equiv\del_\mu \Phi- ig{\tau^a \over 2}A^{a}_{\mu}\Phi.
}\eqn\Ij$$
$v$ is the vacume expectation value of the
Higgs field;
$\bra{0}\Phi\ket{0}=v/\sqrt 2$, by which the gauge boson mass $M_W=M_Z$ is
$gv/2$,
the Higgs
boson mass $M_H$ is ${\sqrt\lambda}v$ and we parametrize the Higgs field as
$${
\Phi={1\over \sqrt 2}\bigl(v+\phi+i \tau^a\chi^a \bigr){0\choose 1}}.
\eqn\Ik
$$
The gauge fixing term $\Lag_{R_{\xi}}$ is given by
$${
\Lag_{R_{\xi}}=(-i)\delta_B \bigg(-\del_\mu \bar C^a A^{a \mu}+{\alpha_g
\over 2}
\bar C^a B^a +\alpha_g M_W \bar C^a \chi^a \bigg)}, \eqn\Il$$
where $\delta_B$ means the BRS transformation
and $\alpha_g$ is the gauge parameter.
The path-integral expression of the $S$-operator is now given in terms of the
following canonical variables;
$${
    \big(\Phi^a,\Pi^{a}_{\Phi} \big) \equiv
 \Bigl\{
    \big(A_{0}^{a},B^a \big),\big(A_i^a,\Pi_A^{ai}\big), \big(B^a,\Pi_{B}^{a}
\big),
    \big(\phi,\Pi_{\phi} \big),\big(\chi^a,\Pi_{\chi}^{a} \big),
    \big( C^a,\Pi_{C}^a \big), \big( {\bar C}^a,\Pi_{\bar C}^a \big)
 \Bigr\}.
    }\eqn\Im$$  \par
Under the spherically symmetric ansatz leading to the sphaleron
as Ratra and Yaffe
\refmark{\ry} introduced,
we obtain the $1+1$ dimensional action of gauge ann Higgs fields given
by them  together with $R_{\xi}$ gauge fixing term. These can be expressed,
following
the prescription by Aoyama, Goldberg and Ryzak,\refmark{\agr} in terms of the
non-contractible loop
parameter $\mu \in [0,\pi]$ and three trial functions $f, h$ and $K$ \foot{
$\Omega(t,r)$ in Ref. 12) is given by $2\mu(t)K(r)$.}.
Here we have chosen that $\mu$ depends only on time $t$ and that $f, h$ and
$K$
are functions of spatial variable $r$.
The neutral Higgs field $\phi_H$ we are concerning is expressed as follows;
$$
  \phi_H ={{2 M_W}\over g}\big[{\rm Re}[{\rm e}^{-i\mu K(r)}(\cos\mu(t)+ih(r)
\sin\mu(t))]-1\big].
\eqn\In$$
After some manipulations, we obtain the reduced euclidean action
with $\mu$
as the dynamical variable;
$$
  S_E[\mu] ={4\pi \over g^2} \int dtdr \big[\half M(t,r)
           {{\dot \mu}}^2+V(t,r){\rm sin}^2\mu \big]
\eqn\Io$$
with $\dot \mu = d\mu/dt$. Here
$$\eqalign{   M(r,t)
    & =4\Bigl(r^2K^{\prime 2}+2(f-K)^2+M_W^2r^2(h-K)^2\Bigr) \cr
    & +4\Bigl(8f(1-f)K(1-K)+M_W^2r^2(1-h^2)(1-K^2)\Bigr){\rm sin}^2\mu,
\cr}$$
$$\eqalign{      V(t,r)
    & =2\Bigl(2f^{\prime 2}+M_W^2r^2h^{\prime 2}+2M_W^2(f-h)^2\Bigr) \cr
    & +\Bigl(8f^2(1-f)^2/r^2+4M_W^2f(1-f)(1-h^2)-4M_W^2f(1-h)^2  \cr
    & +(1/2)M_H^2M_W^2r^2(1-h^2)^2\Bigr){\rm sin}^2\mu \cr}
\eqn\Ip$$
with $f^{\prime}=df/dr$.
Two gauge conditions,
$$\eqalign{
  & h(r){\rm tan}(\mu(t))={\rm tan}(\mu(t)K(r)), \cr
   &\mu K^{\prime\prime}
        +(1/r)\Bigl[  (1/r)\Bigl(f{\rm sin}(2\mu(1-K))
                 -(1-f){\rm sin}(2\mu K)\Bigr)+2\mu K^{\prime}\Bigr]=0,  \cr}
\eqn\Iq
$$
are obtained from the $R_{\xi}$ gauge fixing term. (See Appendix.)
\chapter{Model Calculation}
Let $\rho$ be the shpaleron size.
We choose the trial functions of the following form;
$$
f(r)={\rm tanh}^2(r/\rho),\quad h(r)=K(r)={\rm tanh}(r/\rho),
\eqn\IIa$$
which do reproduce the required behavior of the sphaleron at $r\simeq 0$ and
of the
exponential damping of ${\rm e}^{-M_W r}\sim {\rm e}^{-M_H r}$ as
$r \rightarrow \infty$. In the both regions, fortunately, they also satisfy
the gauge
conditions at $\mu=\pi/2$ (the sphaleron configuration) and
$\mu=0,\pi$ ( the vacua).
These may be satisfactory in views of simplicity of the trial functions.
In terms of
dimensionless time $\tau=M_Wt$ and variational parameter $a=M_W\rho$, the
euclidean
action is written
as
$$
  S_E[\mu] ={4\pi \over g^2} \int d\tau \big[\half M(\mu,a)
           {{\dot \mu}}^2+V(\mu,a){\rm sin}^2\mu \big]
           \eqn\IIb$$
with $\dot \mu = d\mu/d\tau$. Here
$$\eqalign{
     M(\mu,a)&=\Bigl(4(\pi^2/18+4{\rm ln}2-3)a\Bigr)
             +\sin^2 \mu\Bigl((8/5)a+4((\pi^2/18)-(1/3))a^3\Bigr),\cr }$$
$$\eqalign{
     V(\mu,a)&=\Bigl((32/15)/a+2(\pi^2/18+4{\rm ln}2-3)a\Bigr) \cr
             &+\sin^2 \mu\Bigl(8C/a+8(11/5-{\rm ln}2)a
             +\sqrt{2}M_H^2/M_W^2((\pi^2/18)-(1/3))a^3\Bigr) \cr}\eqn\IIc$$
with $C=\int_0^\infty dx(\sinh^4 x/(x^2\cosh^8 x))=0.0916769$,
which are dimensionless mass and potential height respectively. \par
The bounce solution within a self-consistent approximation
is obtained as done in Ref.$\,$21).
We first regard time dependence coming from $\sin^2\mu$ of $M(\mu,a)$ and
$V(\mu,a)$
as weak and replace them by
$M_0(a)$ and $V_0(a)$ respectively for a moment. Then the first integral of
the
equation of motion gives
$$
   (1/2)M_0(a){\dot \mu}^2-V_0(a)\sin^2 \mu = - V_0(a)(1-\k^2),
\eqn\IId$$
where $\k$ is the integral constant such that $0 \leq \k \leq 1$.
The periodic solutions to \IId\ are
$$
     \mu_b(\tau) = \arccos\bigl[ -\k\ \sn(b(\k,a)\tau ;\k)) \bigr]
\eqn\IIe$$
with $b(\k,a)=\sqrt{2V_0(a)/M_0(a)}$, where $\sn(x;\k)$ is the elliptic
function and the period is given by
$$
   b(\k,a) T(E(\k)) = 4nK(\k)  \qquad {\rm with} \quad
                            \qquad n=1,2,3,\cdots \quad ,
\eqn\IIf
$$
where
$K(\k)=\int_0^{\pi/2} d\theta (1-\k^2 \sin^2\theta )^{-1/2}.$
We take the $n=1$ solution without nodes and evaluate the average of
$\sin^2\mu(\tau)$
by it as
$$
  \langle\sin^2 \mu\rangle=\bigl(1+{{\k\sqrt{1-\k^2}/{\arcsin\k}}}\bigr)/2.
\eqn\IIg
$$
By putting this back to $M(\mu,a)$ and $V(\mu,a)$, we
replace them by $M(\k,a)$ and $V(\k,a)$ respectively.
Taking half (one way) of this bounce motion at $E=V(\k=0,a)(1-\k^2)$,
we obtain the tunneling exponent $W(E)/g^2$ by \If\ as
$$
W(E)/g^2=8\pi\sqrt{2M(\k,a)V(\k,a)}\big[E(\k)-(1-\k^2)K(\k)\big]/g^2,
\eqn\IIh
$$
where $E(\k)=\int_0^{\pi/2}d\theta(1-\k^2\sin^2\theta)^{1/2}$.  \par
We remark some physical features of importance. Obviously at $\k=0$, $W(E)=0$
while at $\k=1$, $W(E)$ is minimized at $a=0$ giving $W(E)/g^2=1.027S_{inst}$
where
$S_{inst}=8\pi^2/g^2$. Such a smooth connection between the almost free
passing over
the sphaleron barrier at high energies and the strong instanton suppression
at low
energies is one of natural implications of our formalism.
Given the size $a$, the sphaleron mass is
$$
E_{sph}(a)/M_W={(4\pi/ g^2)} V_0(\k=0,a)
\eqn\IIi$$
as shown in Fig.1, which is minimized at $a=a_0$ and is compared with Yaffe's
numerical result of $E_{sph}^{Y}$\refmark{\Yaffe} in Table I. Such small
deviations
from
the $S_{inst}$ and $E_{sph}^{Y}$
may prove that our trial functions are satisfactory.
Note that the parameter
$a\,({\rm or}\, \rho)$ is not only of the variational character but
discriminates
the symmetry breaking of the electroweak theory.
That is, $a_0\not=0$ to minimize the sphaleron realizes the broken phase
while $a=0$
of the instanton does the symmetric phase as understood from a dimensional
consideration.\refmark{\fot} \par
\centerline{{\bf Fig.1}\quad {\bf Table I}}
Now we evaluate the Fourier transform of the classical Higgs configuration.
The edge
of the bounce is obtained by substituting the solution \IIe\ at the edge back
into
$\phi_{H}$ in \In.
It is given by
$$
 \phi_c^{(in, out)}(r/\rho ) =
  -{{2\sqrt{2\pi}M_W}\over g}{ {\sin^2 \mu_b^{(in, out)}(1-h^2(r/\rho))}
    \over {1+\sqrt{\cos^2\mu_b^{(in,out)}+h^2(r/\rho)\sin^2\mu_b^{(in,
out)}}}},
\eqn\IIj$$
where
$\mu_b^{(in, out)} \equiv \mu_b(\mp T/2) = \arccos(\pm \k).$
Here we have used the first of the gauge conditions \Iq. In calculating the
Fourier
profile,
we approximate the denominator in \IIj\ by a constant $\langle D\rangle$
between
1 and 2 noting that
this is a slowly varying function within this interval. Then we have an
analytic expression;
$$\eqalign{
 \alpha^{(in,out)}({\bf k})/g
 &=\alpha^{*(in,out)}({\bf k})/g    \cr
 &\cong -{ 2 \pi \over g}\sin^2\mu_b^{(in,out)} {{M_W \rho^2}
\big({\vert {\bf k}\vert}^2 + M_H^2 \big)^{1 \over 4}
\over
{\langle D\rangle{\vert {\bf k}\vert}
 \sinh(\pi \rho {\vert{\bf k}\vert}/2)}}\Bigl({\pi \rho {\vert{\bf k}\vert}/2
\over
 \tanh(\pi \rho {\vert{\bf k}\vert}/2)}-1 \Bigr). \cr
}\eqn\IIk$$  \par
%
Equipped with the tunneling exponent $W(E)/g^2$ and the Fourier profile
$\alpha({\bf k}) /g$'s
of the classical configuration, we are ready to calculate the $S$-matrix
element of
$2 \rightarrow n$ Higgs production according to \Ie\  and \Ig. In covariant
normalization, it reads
\hfuzz=5.0pt
$$\eqalign{
   \bra{{\bf k}_1,\cdots,{\bf k}_n} \hat S \ket{{\bf p}_1,{\bf p}_2}
 &=(2\pi)^4
     \delta^{(4)}\bigl( \sum_{i=1,2} p_i - \sum_{j=1}^n k_j \bigr)\cr
    \times {\rm e}^{-W(E)/g^2} {\rm e}^{ -{A^2}(E)/{g^2}}
 &  {{\sqrt{(2\pi)^3 2p_1^0}\alpha^{(in)}({\bf p}_1)
                \sqrt{2\pi)^3 2p_2^0}\alpha^{(in)}({\bf p}_2)}\over{g^2}}
    \prod_{j=1}^n {{\sqrt{(2\pi)^3 2k_j^0}\alpha^{(out)}({\bf
k}_j)}\over{g}}, \cr
}\eqn\IIl$$
where the normalization exponent of the coherent state is given by
$$
{{A^2(E)}\over{g^2}}={1\over{g^2}}
\int_{-\infty}^{\infty}d^3{\bf k}\,\alpha^{(in)}({\bf k})^2
={1\over{g^2}}\int_{-\infty}^{\infty}d^3{\bf k}\,\alpha^{(out)}({\bf k})^2.
$$
\par
The cross section obtained from the above $S$-matrix element in center of
momentum
system at the incident energy $E$ is of the following form.
$$
  \sigma_{2\rightarrow n}=X_0 X_n,
\eqn\IIm$$
where
$$
   X_0={(2\pi)^{10}E^2 \over {4E{\vert{\bf p}\vert}}}{\rm e}^{-2W(E)/g^2}
        {\rm e}^{-2A^2(E)/g^2}\big({{\alpha^{(in)}({\bf p})}\over g}\big)^4
M_W^8 ,
$$
$$
X_n={1\over n!}\int \prod_{j=1}^n \Bigl[{d{\bf k}^3_j\over (2\pi)^3 2k_j^0}
\Bigl({\sqrt{(2\pi)^3 2k_j^0}\alpha^{(out)}({\bf k}_j) \over g}\Bigr)^2\Bigr]
\delta(\sum_{j=1}^n k_j^0-E) \delta^{(3)}(\sum_{j=1}^n{\bf k_j})
\eqn\IIn$$
with ${\vert {\bf p}\vert}=\sqrt{(E/2)^2-M_H^2}$ and
$k_j^0=\sqrt{{\bf k}^2+M_H^2}$.
\chapter{Numerical analysis---suppression versus enhancement}\par
%
The $n$-independent part $X_0$ always suppresses
$\sigma_{2\rightarrow n}$ as strongly as the instanton does. At low energies,
the tunneling exponent $2W(E \simeq 0)/g^2 \simeq 2S_{inst}$ at $a=0$ leads
to
this suppression as remarked before.
At high energies with ${\vert{\bf p}\vert}\simeq E/2$ and $E\simeq E_{sph}$,
the
tunneling
suppression does not work while the incident profiles damp as
$$
\big(\alpha^{(in)}({\bf p})\big)^{4} \equiv
\alpha^{(4)}_{(in)}({\bf p})\sim {\rm e}^{-\pi \rho E_{sph}}={\rm e}^
{-a (\pi E_{sph}/M_W)}.
\eqn\IIIa$$
Referring to Table I, we see that  $(\pi E_{sph}/M_W)$ is comparable to or
even larger
than $2S_{inst}$. The physical reason for this high energy suppression is
that
the Fourier profile of the sphaleron as an extended object damps for large
$\vert {\bf p}\vert$ more
rapidly than any power of $\vert {\bf p}\vert$.\par
Hereafter we fix the
gauge boson mass as $M_W=80.6$
GeV
and $SU(2)$ gauge coupling $g^2=0.418$. Then the sphaleron
mass in Table I is $E_{sph}$=8.053, 9.845, 10.93 and 11.95 TeV for $M_H/M_W$=
0, 1.0, 1.5
and 2.0
respectively. We also fixed $\langle D\rangle$=1.0 in the denominator of
\IIk, though
this may
somewhat over-estimate $\alpha^{(in, out)}$. \par
%
The next task is to evaluate how the multiple of profiles and the phase space
volume of
the
outgoing $n$ particles compete with the suppression factor.
As $X_n$ has no axial symmetry, let us first imagine that the Higgs particles
would
be produced isotropically provided that $n$ is large enough, so that the
angular part
of
the ${\bf k}$ integration could be done. Then a single outgoing particle
contributes
as
$$
      \int d^3{\bf k}\Bigl({{\alpha^{(out)}({\bf k})}\over
g}\Bigr)^2\rightarrow
      \int dk{{\alpha^{(2)}_{(out)}(k)}\over{ g^2}}
      \quad{\rm with}\quad
      \alpha^{(2)}_{(out)}(k)=4\pi k^2(\alpha^{(out)}({\bf k}))^2.
\eqn\IIIb$$
$\alpha^{(2)}_{(out)}(k=0)=0$ independently of $\mu$ and $a$ while it damps
rapidly
as ${\rm e}^{-\pi\rho k}$ for $k \gg M_W$,
so that $\alpha^{(2)}_{(out)}(k)$ always has a sharp peak at $k\rho \sim$
O(1) corresponding to the sphaleron size with the width of O(1/$M_W$).
%
%
An example is given in Fig.2. Let us denote this peak position as $k^{*}$ and
the number of the corresponding produced particles as $n^{*}\simeq
E/\sqrt{k^{* 2}+M_H^2}$. These together with the isotropic distribution would
provide the saddle point of $X_n$ satisfying the energy-momentum
conservation, so that
we may make an approximation for $n\gg 1$;
$$
X_n \rightarrow \bar X_n=(1/n!)(M_W\alpha^{(2)}_{(out)}(k^{*})/g^2)^n/M_W^4.
\eqn\IIIc$$
{}From this we obtain $\sigma_{2\rightarrow n}$ by \IIm\ and $\sigma_{tot}$
by
summing it to the maximum possible $n$.  \par
In order to have some notions how the enhancement by $X_n$ struggles against
the suppresion by
$X_0$, we make a step further to obtain $\sigma_{tot}$  by exponentiation;
$$ \sigma_{tot} = \sum_n \sigma_{2\rightarrow n}\sim {1\over{M_W^2}}X_0
          \times  {\rm e}^{M_W\alpha^{(2)}_{(out)}(k^{*})/g^2}.
\eqn\IIId$$
We should say that this would give an over estimate of the enhancement factor
as
$n$ may be rather limited around $n^{*}$.\par
\centerline{{\bf Fig.2}}
%
Fig.3 shows the typical pattern of the sphaleron deformation. At the point
$a=a_0$
that minimizes $E_{sph}$, $\sigma_{tot}(a_0)$ there does not give the maximum
of
$\sigma_{tot}$, but $\sigma_{tot}$ has two local maxima: one
$\sigma_{tot}(a^{(+)})$ at
$a^{(+)}>a_0$ and the other $\sigma_{tot}(a^{(-)})$ at $a^{(-)}<a_0$
respectively.
The mechanism to give $\sigma_{tot}(a^{(\pm)})$ can be traced back to
$\alpha^{(in,out)}\propto a^2$ in \IIk\ . As $a$ increases from $a_0$, the
barrier
becomes higher and the
sphaleron gets fatter indeed, but the $n$-multiple of
$\alpha^{(2)}_{(out)}\propto a^4$
gives the larger effect to increase $\sigma_{tot}$. On the other hand, at
$a^{(-)}\ll 1$,
all the effects coming from $\alpha^{(in, out)}$ vanish while
$2W(E)/g^2$ in this symmetric phase coincides with $2S_{inst}$,
so that $\sigma_{tot}(a^{(-)})$ is controlled as
$(1/M_W^2){\rm e}^{-2S_{inst}}$ which is numerically larger than
$\sigma_{tot}(a_0)$ .     \par
\centerline{{\bf Fig.3}}
Fig.4 shows energy dependence of $\sigma_{tot}(a^{(\pm)})$ and
$\sigma_{tot}(a_0)$.
In the case of
$M_H/M_W=1.0$, $\sigma_{tot}(a^{(-)})$ of the instanton suppression should
turn
into $\sigma_{tot}(a^{(+)})$ as $E$ increases. The reason why
$\sigma_{tot}(a^{(+)})$
is smaller than $\sigma_{tot}(a^{(-)})$ for $M_H/M_W > 1.2$ will be explained
in the next section. Anyhow, $\sigma_{tot}$ are
all lower than the unitarity bound $\sigma_{unitarity}=16\pi/E^2$. \par
\centerline{{\bf Fig.4}}
\chapter{Concluding remarks}\par
1. The structure of the $\Delta B \not= 0$ scattering is quite clear in our
formalism based on path-integral expression of the
time-evolution operator during tunneling at non-zero energies, which is
dominated by
the classical bounce solution
giving the tunneling exponent that bridges from the instanton to the
sphaleron.
At low energies, the exponent gives the instanton suppression as usual.
The Fourier profile
$\alpha^{(in,out)}({\bf k})/g$ at the edges of the tunneling
may correspond to the residue of Green's function
in the instanton background in the LSZ formalism, but the former damps
rapidly for
large momenta since the
sphaleron is a classical lump of field with a finite size $\sim 1/M_W$.
At high energies the overlap of the incident wave with the entrance of
tunneling
leads to such a suppression comparable with that
given by the instanton.
Importance of overlap of the incident
waves with the entrance of tunneling configuration
at finite energies, which could keep $\sigma_{tot}$ to be very small, was
also
stressed by
Banks {\it et al}..\refmark{\bfds} \par
2. The multiple of the outgoing profiles together with the phase space volume
indeed
enhances
$\sigma_{tot}$, but,
in so far as the maximum number of produced particles is $n\sim 100$, the
multiple may be
too weak to overcome
the suppression at high energies.
We should mention that in the case of extremely
large number of produced particles,
the enhancement factors exceed the instanton-like suppression and do
not prevent
$\sigma_{tot}$ from a rapid increase. We show two examples. \nextline
(i) $\sigma_{tot}(a^{(+)})$ at $E=E_{sph}$ in Fig.4 sharply depends on
the maximum number $n$ through
$M_H/M_W$. If $M_H/M_W \sim 0.5$ fictitiously so that $n \sim 200$, an
extrapolation
of $\sigma_{tot}(a^{(+)})$ would surely reach $\sigma_{unitarity}$.
For $M_H/M_W > 1.2$, in contrast, $n$ is too small to overcome the
suppression factor.
\nextline
(ii) In the $O(3)$ model in I, $\sigma_{tot} \sim 10^{-30} $ for
$g^2 \sim 0.1 (n \sim 100)$
while $\sigma_{tot} > 1 $ (unitarity bound in $1+1$ dimensions) for
$g^2 < 0.022 (n > 362)$.
\nextline
In other words, $g^2=0.418$ of the $SU(2)$ gauge coupling constant may be too
large
to take the weak coupling limit, or, realistic $n\sim100$ may be dangerous to
make
an easy exponentiation etc. \par
One might claim that any $\sigma_{tot}$ should reproduce the one by the
instanton
calculus at low energies,\refmark{\KRT} given by\refmark{\Mattis}
$
\sigma_{tot}\sim {\rm exp}\Bigl[{({16\pi^2}/ g^2)}[-1 + {\rm const}.\times
(E/E_{sph})^2]\Bigr].
$
We feel, however, that the weak coupling approximation used there would be
inadequate
as just mentioned. \par
3. The strong dependence of $\sigma_{tot}$ on $n$ may be
related to how drastically a severe cancellation
takes place to give
$\sigma_{tot}$ at high energies as shown in Fig. 3.
Because of the severe and delicate cancellation together with possible
corrections
in the next-to-leading order,\foot{We might expect that the corrections would
not
be significant in our cases, since the leading order result itself is much
smaller
than the
unitarity bound.}
we do
not asert here that $\sigma_{tot}$ with $\Delta B \not=0$ is actually
extremely small
as shown in
Fig.4.\foot{We have also to take the gauge boson production into account, as
the
sphaleron would decay dominantly into them\refmark{\zad}.}
It seems unlikely, however, that it grows as largrly as to reach the
unitarity
bound at energies $E \le E_{sph}$.
An exponential suppression of $\sigma_{tot}$ at high energies was strongly
suggesed
by Shaposhnikov.\refmark{\shap} \par
4. We finally repeat a remark in I that unitarity is a subtle issue in the
$\Delta B\not=0$ high
energy scattering.
In any formalism the scattering is considered to be a transition
between states each constructed on minima of classical energy functional
space and these (quasi-stationary) states belong to the eigenstate of $B$.
But the
hamiltonian here does not commute with $B$ because of the chiral
anomaly. Hence the asymptotic states, which belong to the eigenstate of
hamiltonian in
usual scattering
problem, are not the eigenstate of the hamiltonian. So the $\Delta B\not=0$
process
corresponds
to off-diagonal $S$-matrix elements. We have tacitly assumed that
these elements are small compared with the diagonal ones, which would be
consistent with
our $\sigma_{tot}$ far below the unitarity bound. If $\sigma_{tot}$ with
$\Delta B \not=0$
were to reach the unitarity bound, one can
not rely on any formalism available at present but has to develop a basically
new
formalism to treat the $\Delta B\not=0$ scattering.
\vskip 1.0cm
\centerline{\bf Acknowledgement}
The authors would like
to express their cordial gratitude to their colleagues of Saga, Kyushu, and
Kinki Universities for discussions and encouragement. S.O. is supported by
Grant-in-Aid of the Ministry of Education, Science and Culture(No. 03640276).
\vskip 1.0cm
\centerline{\bf Appendix: Gauge fixing conditions in $R_{\xi}$ gauge} \par
After performing the BRS transformation, the original gauge fixing term
\Il\ in the text is expressed as
$$\eqalign{
\Lag_{R_{\xi}}&=-\del_\mu B^a A^{a \mu}+\alpha_g M_W B^a \chi^a + {\alpha_g
\over 2}
B^{a} B^{a} - i \del_\mu \bar C^a D^\mu C^a    \cr
&+
i \bar C^a \big[\alpha_g {M_W}^2 \delta^{ac}+{\alpha_g \over 2}gM_W(\phi
\delta^{ac} +
\epsilon_{abc}\chi^b) \big]C^c.  \cr
}\quad{({\rm A}.1)}$$
Let us introduce the spherical symmetric ansatz following Ratra and Yaffe;
\refmark{\ry}
$$\eqalign{
A_0 &\equiv A^{a}_{0}{\tau^a \over 2}={1 \over 2g}a_0(t,r) \tau^a \hat x^a ,
\cr
A^j &\equiv A^{j a}{\tau^a \over 2}={1 \over 2g}\big[{\alpha(t,r) \over r}
{\rm e}^{j(1)}
+ {{1+\beta(t,r)}\over r}{\rm e}^{j(2)}
 + a_1(t,r){\rm e}^{j(3)} \big],  \cr
\Phi&={1 \over g}\big( \sigma(t,r) + i \eta(t,r) \tau^a \hat x^a \big){0
\choose 1}
\cr
}\quad{({\rm A}.2)}$$
with
$$
{\rm e}^{j(1)}\equiv \tau^j - \tau^a \hat x^a \hat x^j,  \quad {\rm e}^{j(2)}
\equiv
i(\tau^a \hat x^a \tau^j - \hat x^j)  \quad {\rm and}\quad
{\rm e}^{j(3)}\equiv \tau^a \hat x^a \hat x^j .
$$
After dealing with the subsidiary fields in the standerd way,
$\Lag_{R_{\xi}}$ is written as
$$
\Lag_{R_{\xi}}={v \over \sqrt2 g}\big(a_1\eta'+2{\alpha \eta \over r^2}\big)
          -\alpha_g{v^2 \over 4}{\eta}^2 - {1 \over \alpha_g}{1 \over
2g^2}Y^2 ,
\qquad{({\rm A}.3)}
$$
where $Y$ is given by
$$
Y \equiv \dot a_0 - {a_1}' + {2 \over r^2} \alpha - {2 \over r} a_1 .
\qquad \quad{({\rm A}.4)}
$$
We put $\Lag_{R_{\xi}}=0$, which leads to two gauge fixing conditions,
$$
\eta(t,r)=0 \quad {\rm and} \quad Y(t,r)=0.
\quad{({\rm A}.5)}$$
They are rewritten in terms of $\mu(t), f(r), h(r)$ and $K(r)$ as \Iq\ in the
text. Note that $\eta(t,r)=0$ implies vanishing of the imaginary part of
$\phi_H$ in
\In .

\par \penalty-400 \vskip\chapterskip
   \spacecheck\referenceminspace \immediate\closeout\referencewrite
   \referenceopenfalse
   \line{\fourteenbf\hfil REFERENCES\hfil}\vskip\headskip
   \input reference.aux
   
\vskip3cm
\noindent
{\bf Table I.} The minimized $E_{sph}(a_0)$ by \IIi. $r$ denotes the ratio of
$E_{sph}(a_0)$ to $E_{sph}^{Y}$ by
Yaffe \refmark\Yaffe obtained by minimizing the energy functional.
\vskip0.3cm
\begintable
 $M_H/M_W$ | $E_{sph}/(M_W/g^2)$ | $a_0$ | $r$    \cr
 0.0       | 41.76               | 1.742 | 1.09   \cr
 1.0       | 51.06               | 1.150 | 1.12   \cr
 1.5       | 56.71               | 0.938 | 1.18   \cr
 2.0       | 62.00               | 0.874 | 1.25   \endtable
\endpage
\vskip2.5cm
{\centerline   {\bf Figure Captions}}\par
\noindent
{\bf Fig.1.} $E_{sph}(a)$ {\it vs.} $a$. The minimized point $a=a_0$ is shown
by
the arrow.\par
\noindent
{\bf Fig.2.} $\alpha_{(out)}^{(2)}(k)$ {\it vs.} $k$ which is peaked at
$k^\ast$ (denoted by the dotted line) for $M_H/M_W=1.0$ and $a=a_0=1.150$,
where
$E_{sph}=9.845$ TeV.  \par
\noindent
{\bf Fig.3.} ln$\sigma_{tot}$ (solid curve) by \IIId\ {\it vs.} $a$  for
$M_H/M_W$=1.0 and at
$E$=9.5 TeV. The relevant sphaleron mass is $E_{sph}(a_0)=$9.845 TeV. Note
that ln
$\sigma_{tot}$ is given by the difference between two large quantities,
ln$(M_W^2 X_0)$ and $M_W\alpha_{(out)}^{(2)}(k^\ast)/g^2$ (the both denoted
by broken
curves). Three quantities contributing to the former (dotted curves) are
shown. The
corresponding $k^\ast$ and $n^\ast$ explained in the text are also shown.\par
\noindent
{\bf Fig.4.} $\sigma_{tot}(a)$ obtained by making use of \IIIc.
$\sigma_{tot}(a_0)$ (dotted curve) is $\sigma_{tot}$ in the case when
the the sphaleron does not deform. $\sigma_{tot}(a^{(+)})$ (solid curve) and
$\sigma_{tot}(a^{(-)})$ (broken curve), the latter being almost independent
of
$M_H/M_W$, are the local maxima.
These quantities are plotted {\it vs.} $E$ for $0\le E \le E_{sph}$,
$E_{sph}$
being given in {\bf Table I}.
\end